\documentstyle[prb,epsfig,aps]{revtex}
\input epsf

\newcommand{\centps}[2]{
        \begin{center}
                \epsfig{file=#1,height=#2mm}
        \end{center}
}

\newcommand{\twofig}[3]{
        \begin{center}
                \epsfig{file=#1,height=#3truecm}
                \epsfig{file=#2,height=#3truecm}
        \end{center}
}

\newcommand{\threefig}[4]{
       \begin{center}
                \epsfig{file=#1,height=#4truecm}
                \epsfig{file=#2,height=#4truecm}
                \epsfig{file=#3,height=#4truecm}
       \end{center}
}

\newcommand{\fourfig}[5]{
        \begin{center}
            \mbox{
                \setlength{\epsfxsize}{#5truecm}
                \setlength{\epsfysize}{#5truecm}
                \epsfbox{#1}
                \setlength{\epsfxsize}{#5truecm}
                \setlength{\epsfysize}{#5truecm}
                \epsfbox{#2}}
           \mbox{
                \setlength{\epsfxsize}{#5truecm}
                \setlength{\epsfysize}{#5truecm}
                \epsfbox{#3}
                \setlength{\epsfxsize}{#5truecm}
                \setlength{\epsfysize}{#5truecm}
                \epsfbox{#4}}
        \end{center}
}

\begin{document}
\draft

\twocolumn[\hsize\textwidth\columnwidth\hsize\csname @twocolumnfalse\endcsname

\title{Supercurrent switching in Three- and Four- Terminal
Josephson Junctions}

\author{H. Tolga Ilhan and Philip F. Bagwell \\
Purdue University, School of Electrical Engineering \\ 
West Lafayette, Indiana 47907}

\date{\today}
\maketitle

\begin{abstract}

Control of the Josephson current by varying a gate current has
recently been demonstrated in both 4-terminal and 3-terminal
junctions. We show that, when the the gates are weakly coupled to the
Josephson junction, the Josephson current versus gate current (or
versus gate voltage) relation is the same for both the 4- and 3-
terminal geometries.  At low temperature, the supercurrent switches
abruptly as a function of the gate voltage, but only slowly as a
function of the gate current.

\end{abstract}

\pacs{PACS numbers: 74.80Fp, 74.50+r, 73.20.Dx}

submitted to Journal of Applied Physics

] \narrowtext
\section{Introduction}
\indent

Transistors with superconducting sources and drains have been
fabricated over the last 10 years, using field effect control of the
supercurrent flow.~\cite{kleins} For the field effect to change
carrier density typically requires a gate voltage of the order of
volts, while superconducting energy gaps ($E_{\rm gap} = 2\Delta$) are
at best of order millivolts. To overcome this incompatibility in
voltage scales, van Houten suggested using quantum
confinement~\cite{vanhout} in the channel to incorporate a new small
energy scale into the transistor. Due to the added confinement energy,
a small change in gate voltage can produce a large change in the
supercurrent flow.

A small energy scale occurs naturally in the superconducting
transistor, even without the added quantum confinement. The two
superconductors themselves form a special type of `barrier' for
electrons, where the electrons `Andreev reflect' for the
superconductor as a hole.~\cite{andreev,datta} Repeated Andreev
reflections from the two superconductors form a closed path, leading
to energy level quantization inside the device channel. These `Andreev
energy levels' form inside the superconducting gap, with more bound
levels forming as the channel length increases (just as in the
standard type of quantum well). A large fraction of the Josephson
current flows through the Andreev levels~\cite{been1,vwees,bagwell},
where a quasi-particle moving once around this closed path corresponds
to the transfer of a pair of electrons across the device channel.

Figure~\ref{fig:geom} shows a type of gated Josephson junctions used
in two different recent experiments~\cite{morpurgo,schalpers} to
control the Josephson effect.  The gate terminal are all normal metals
(N$_1$,N$_2$) held at voltages ($V_1$,$V_2$). The superconductors (S)
are grounded.  Electrons can tunnel from the gates into the normal
region (N) of the Josephson junction with probabilities
($T_1,T_2$). Setting the tunneling probability $T_2=0$ decouples one
of the gates from the Josephson junction, forming a three terminal
junction.

Morpurgo et al.~\cite{morpurgo} have recently used current flow
through the two opposing gates to switch off the Josephson current in
a 4-terminal junction. Control of the Josephson current with a gate
voltage (or current) in a more conventional 3-terminal geometry has
also been shown by Sch\"{a}lpers~\cite{schalpers} et al..  Both recent
experiments are based on the mechanism of van Wees et al.~\cite{vwees}
that, because the superconductor cannot inject electrons into the
bound Andreev levels, the electrochemical potential $\mu_B$ of the
bound levels will float to the probe voltage (e.g. $\mu_B = eV_1$ in a
3-terminal geometry) rather than to the electrochemical potential of
the superconductors ($\mu_S \equiv
0$). Refs.~\onlinecite{volkov}-\onlinecite{wilhelm} have also used
this principle to analyze the the current flow in different types of
gated Josephson junctions.

\begin{figure}[htb]
\centps{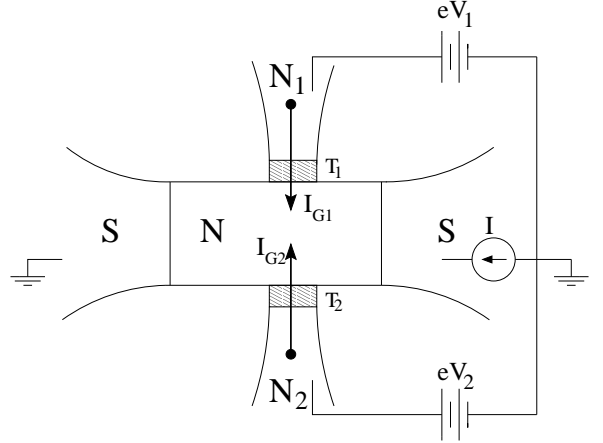}{60}
\caption{ A four terminal Josephson junction. The gate currents 
$I_{\rm G1}$ and $I_{\rm G2}$
passing through the normal region control the supercurrent $I$ by
altering the occupation of bound energy levels in the normal region.}
\label{fig:geom}
\end{figure} 

\section{Four Terminal Junction}
\indent

\begin{figure}[htb]
\twofig{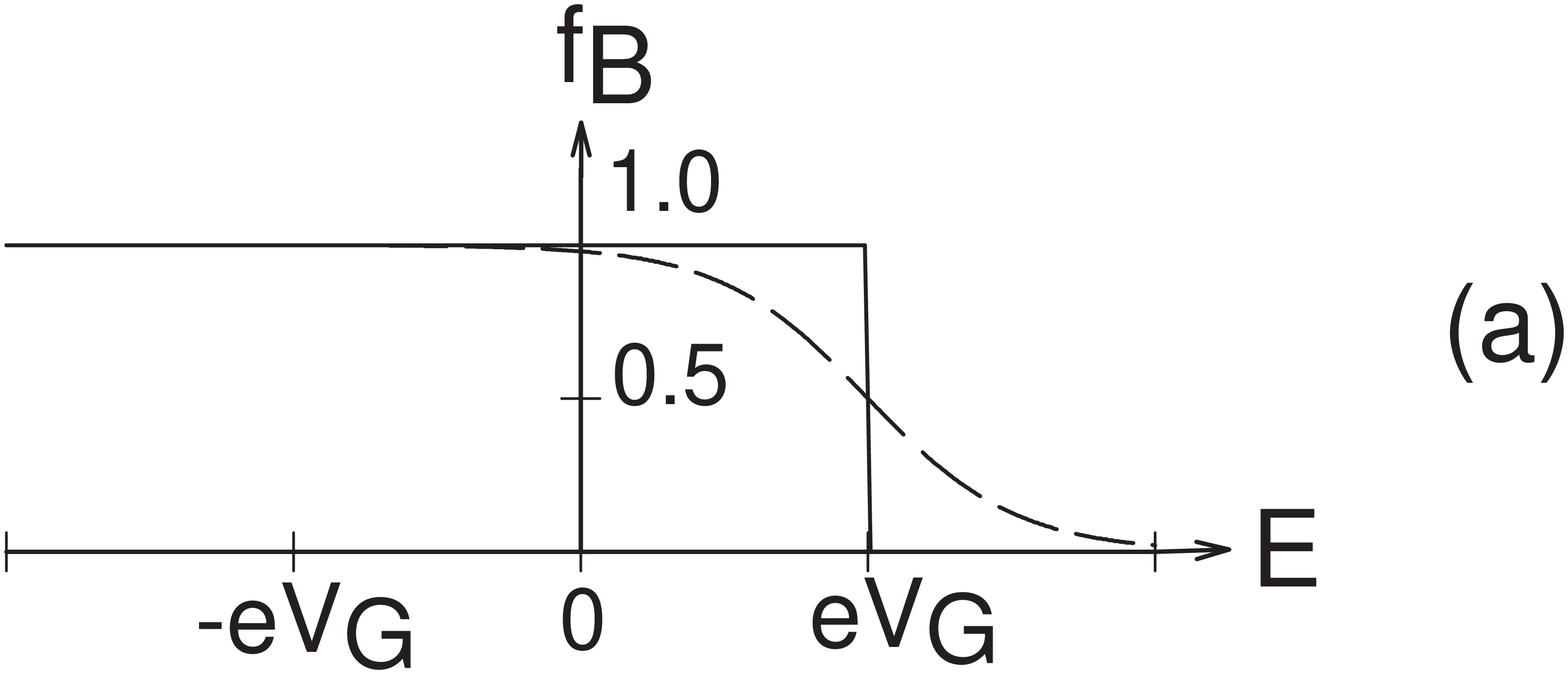}{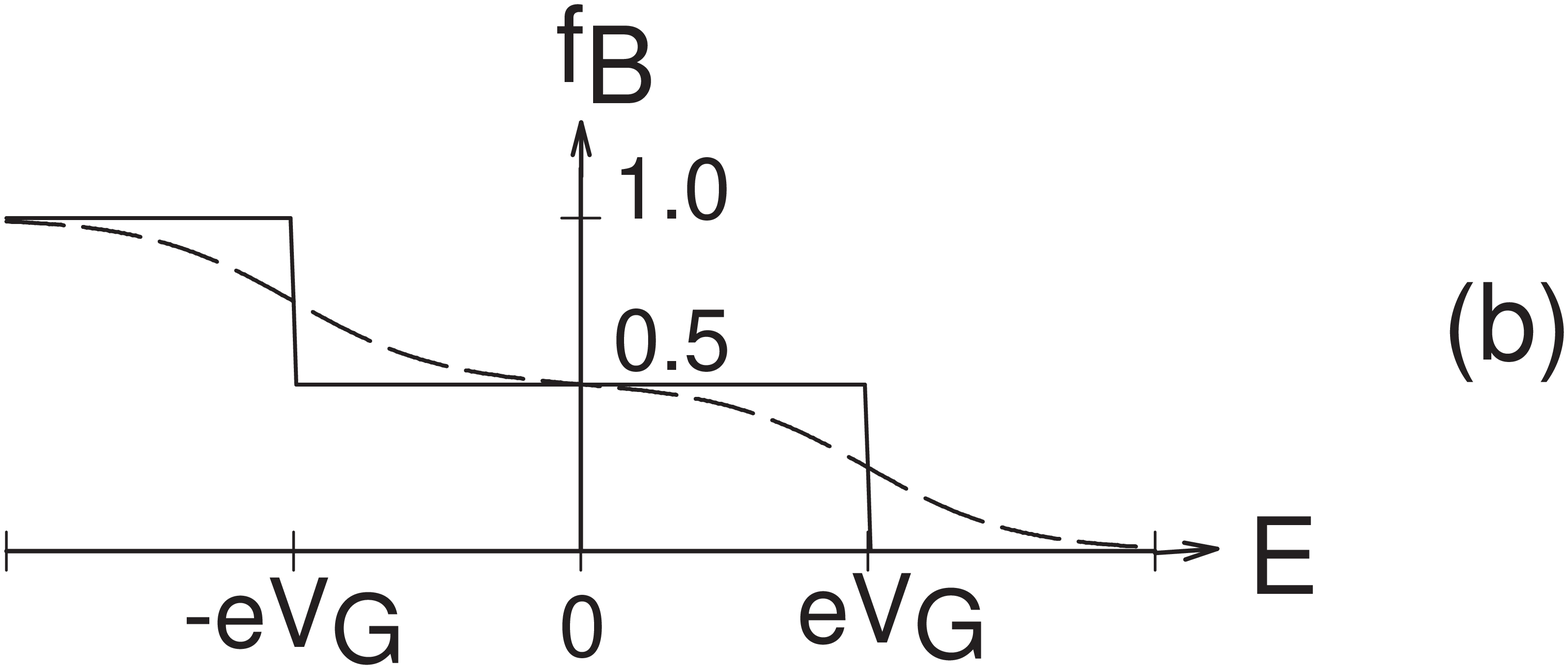}{4.0}
\caption{Occupation factor $f_B(E)$ of the supercurrent carrying
energy levels ($|E| < \Delta$) versus electron energy (referenced to
the Fermi level of the superconductors $\mu_S = 0$) in the (a) three
terminal and (b) four terminal Josephson junctions. Both occupation
factors give the same terminal currents, even though the occupation
factor in (b) is clearly not an equilibrium Fermi factor. Heating the
distributions (dotted lines) smears the switching of the Josephson
current, so that the device performs better with cold electrons
(solid).}
\label{fig:fermi}
\end{figure}

Consider next the Andreev level occupation in a 4- terminal Josephson
junction. The occupation factor $f_B$ for the Andreev levels (inside
the normal region (N) of the Josephson junction where the Andreev
levels exist) is now given by
\begin{equation}
f_B = \frac{T_1 f_1 + T_2 f_2}{T_1 + T_2} .
\label{andfermi}
\end{equation}
The occupation factors $f_1$ and $f_2$ in Eq.~(\ref{andfermi}) are the
Fermi factors of the two gates, namely $f_1 = f(E - e V_1)$ and $f_1 =
f(E - e V_2)$. Here $V_1$ and $V_2$ are the gate voltages and $f(E) =
1/ [1 + \exp(E/k_BT)]$ is the equilibrium Fermi
factor. Equation~(\ref{andfermi}) states that the occupation factor
for an electron in the normal region is simply the probability that
the electrode originated from probe $i=1,2$ times the occupation
factor $f_i$ of probe $i$. The occupation factor in
Eq.~(\ref{andfermi}) is the same as for a small metallic grain,
semiconductor quantum dot, etc. connected to two normal metal
leads. For the 3-terminal device ($T_2 = 0$), Eq.~(\ref{andfermi})
gives the occupation factor of the bound levels is simply the Fermi
factor of the gate $f_B = f_1$. The bound level occupation factors in
the 3- and 4- terminal devices are shown in Fig.~\ref{fig:fermi}.

The Josephson current flow through the either device in
Fig.~\ref{fig:geom} is divided into two different energy
regions~\cite{bagwell} as $I = I_B + I_C$. The `continuum' current
$I_C$ flows at energies outside the superconducting gap ($|E| \ge
\Delta$), and is essentially not affected by the additional gates
(neglecting gate leakage).~\cite{chang} The current $I_C$ is therefore
essentially the same for both 3-terminal and 4-terminal devices. The
`bound level' current $I_B$ flows in the energy range within
$|\Delta|$ of the Fermi level ($|E| \le \Delta$), and is given
by~\cite{chang}
\begin{equation}
I_B = \sum_n I_n(E_n) f_B(E_n) \; .
\label{bcurr}
\end{equation}
Here $E_n$ is the energy of the Andreev level and $I_n$ is the
current carried by the level. In the weak coupling limit ($T_1 \ll 1$,
$T_2 \ll 1$), where the wavefunctions inside the normal region are
essentially unchanged by adding the gates, $I_n$ is the current
carried by bound levels in the Josephson junction before adding the
gates.  Inserting Eq.~(\ref{andfermi}) into Eq.~(\ref{bcurr}) gives
\begin{eqnarray}
I_B = \frac{T_1}{T_1 + T_2} \sum_n I_n(E_n) f(E_n - e V_1) 
\nonumber \\
+ \frac{T_2}{T_1 + T_2} \sum_n I_n(E_n) f(E_n - e V_2) \; .
\label{34connection}
\end{eqnarray}
Equations~(\ref{andfermi})-(\ref{34connection}) are easily
generalized to any number of gates connected to the Josephson
junction.

Morpurgo et al. used a symmetrical device (for which we define the
gate coupling $T_1 = T_2 \equiv \epsilon$), passing a gate current
$I_{G1} = - I_{G2}$, so that the gate voltages satisfy $V_1 = -
V_2$. One can easily verify~\cite{chang} that the bound level current
$I_B$ does not depend on the sign of the gate voltage in a 3-terminal
structure, so in this limit Eq.~(\ref{34connection}) reduces to
\begin{equation}
I_B^{(\rm 1-gate)} = I_B^{(\rm 2-gates)} \; .
\end{equation}
To plot the Josephson current versus gate current (for the symmetrical
4-terminal device in the weak coupling limit), we can therefore use
the same formalism previously developed~\cite{chang,ilhan} for the
3-terminal junction. Specifically, we use the one-dimensional
Josephson junction model of Ref.~\onlinecite{ilhan}, where the normal
region extends over $0<x<L$ and includes a tunnel barrier having
transmission probability $T$ at $x=a$. Ref.~\onlinecite{ilhan} uses
the simplest type of scattering matrix to describe the connection of
the gate to the Josephson junction, which neglects backscattering of
quasi-particles due to the gate.  Additional physics may develop if
one allows the gate to alter the quasi-particle trajectories through
backscattering.~\cite{wendinan}

\twocolumn[\hsize\textwidth\columnwidth\hsize\csname @twocolumnfalse\endcsname
\begin{figure}
\fourfig{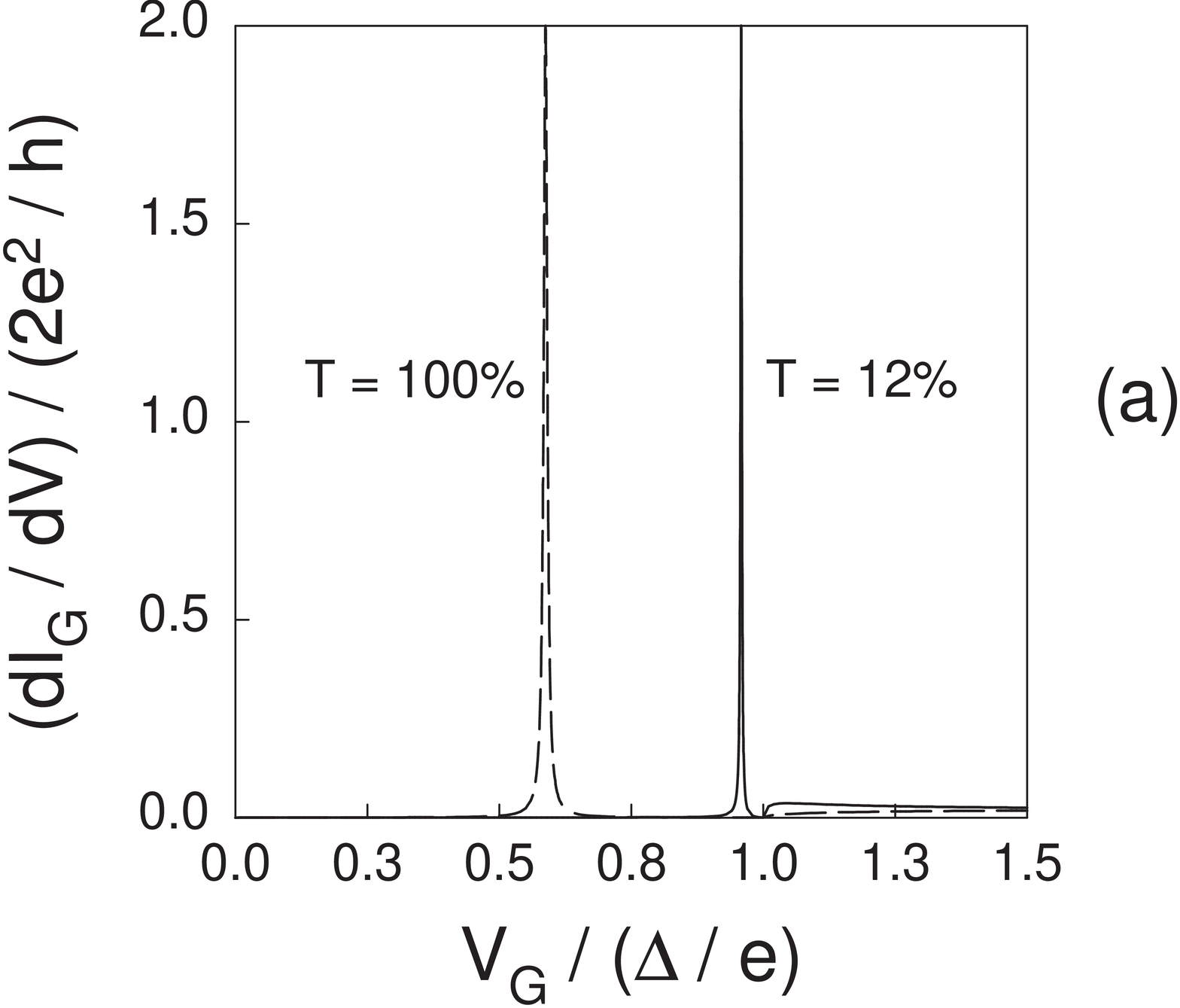}{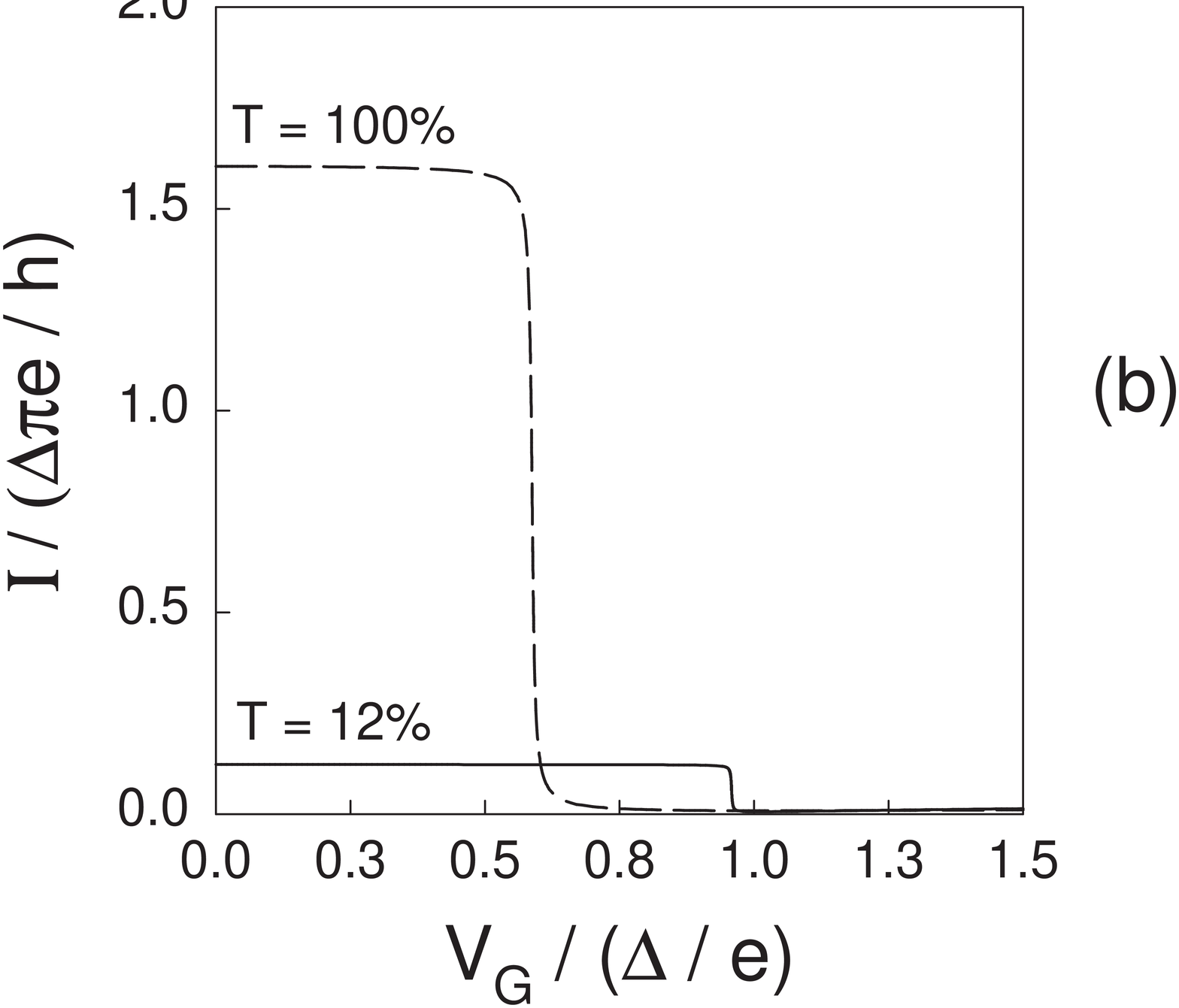}{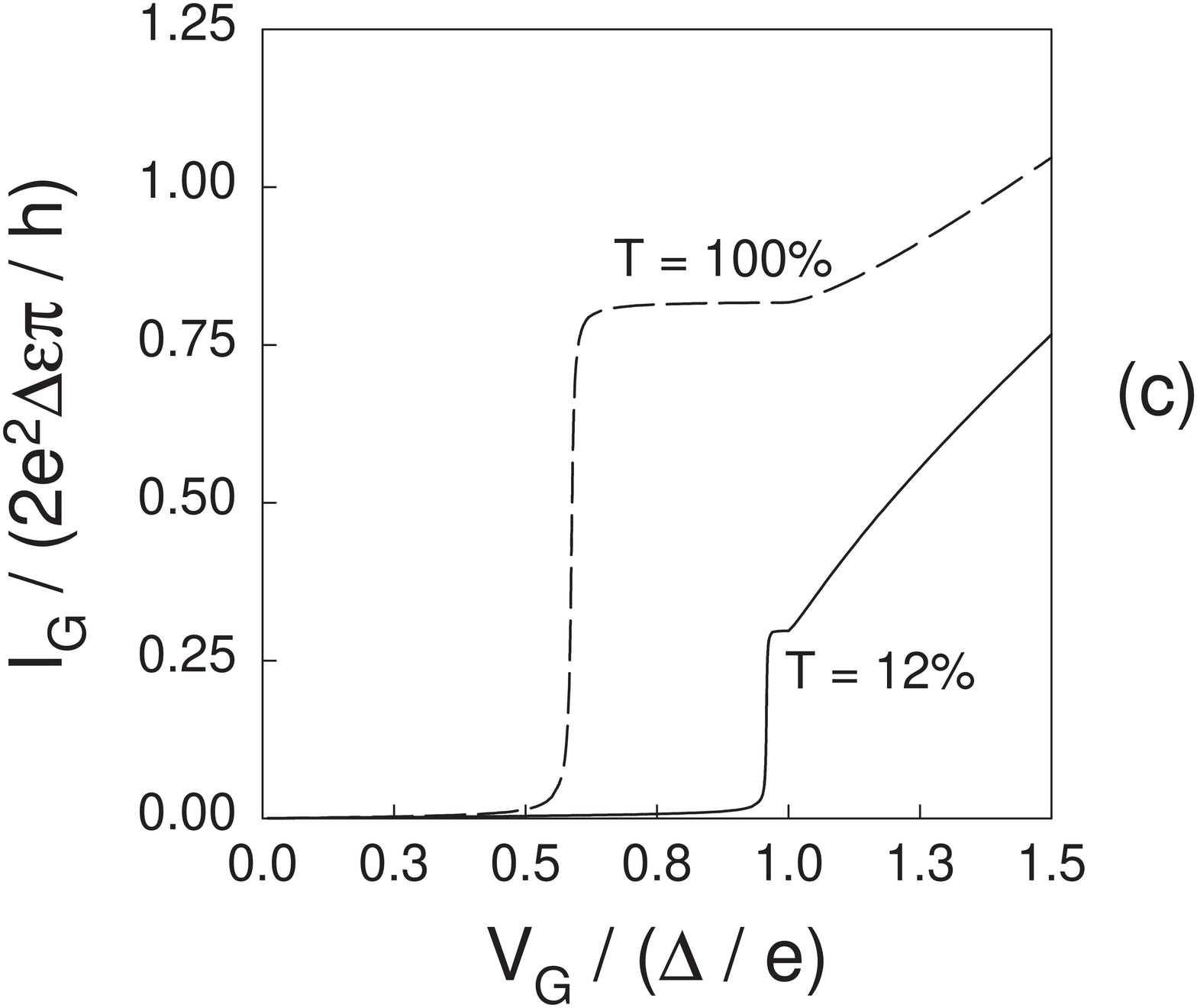}{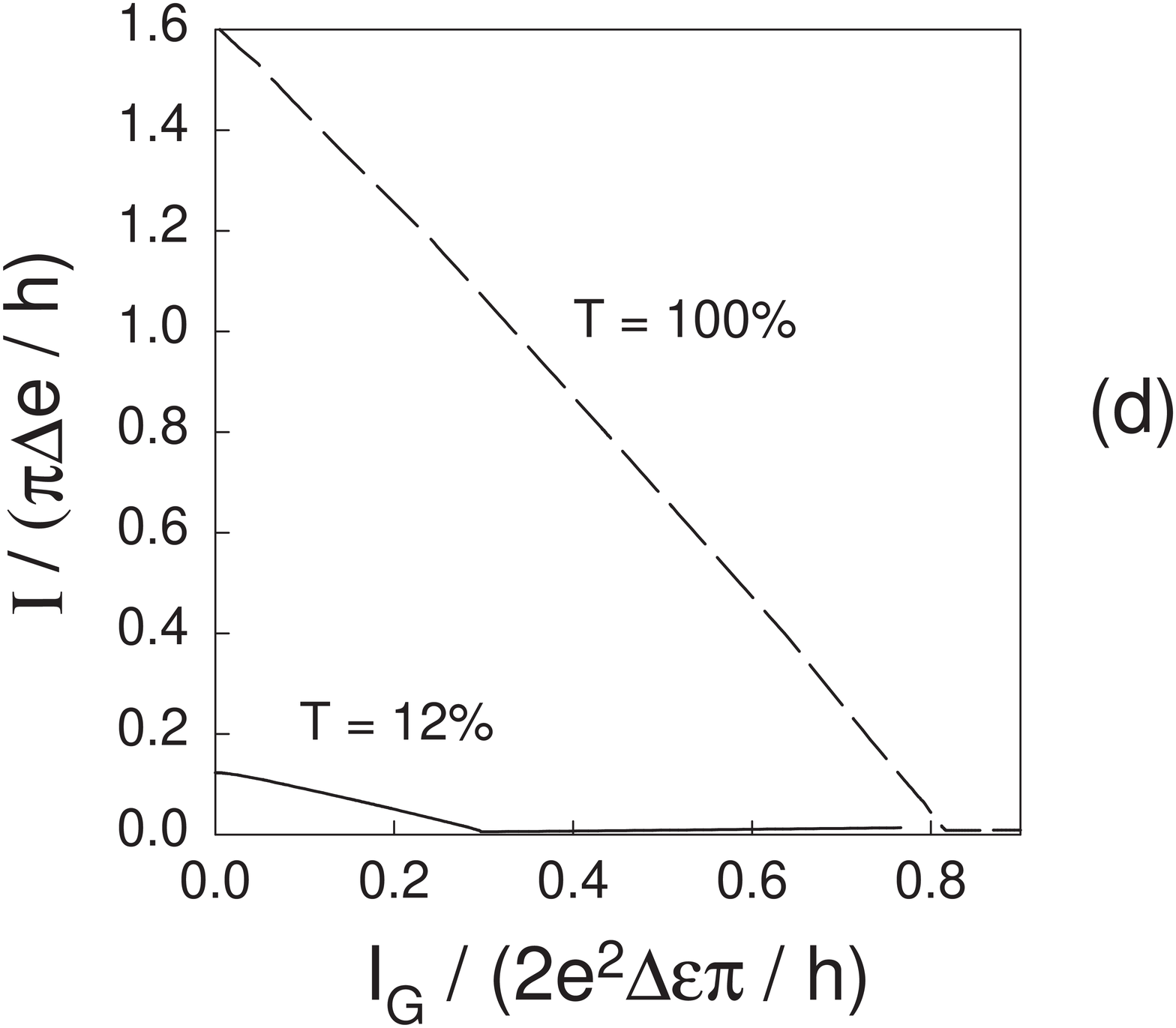}{6.5}
\caption{Terminal characteristics for the short ($L \ll \xi_0$)
Josephson junction. (a) Differential conductance along the gate
versus gate voltage. (b) Josephson current versus gate voltage.
(c) Gate current versus gate voltage. (d) Josephson current versus gate
current. The paramater $T$ is the probability to transmit across
the normal region.}
\label{fig:shortJJ}
\end{figure} 
] \narrowtext

Even though the 4-terminal device has a `double step' Fermi
distribution, shown in Fig.~\ref{fig:fermi}(b), this unusual
nonequilibrium distribution function is irrelevent to the Josephson
current switching. The current is the same as if the bound levels had
the standard equilibrium Fermi distribution of
Fig.~\ref{fig:fermi}(a). The temperature of the electrons in this
distribution, whether they are `hot' or `cold', is also not important
to the switching effect. Switching of the Josephson current versus
gate voltage or gate current will be sharper if the electrons remain
`cold'. For the 3-terminal device having the distribution function of
Fig.~\ref{fig:fermi}(a), there is also a voltage drop of $V_G$ across
the NS junctions. The total voltage drop between the two
superconductors is still zero, as the voltage difference across the
left NS junction is the negative of the voltage drop across the right
NS junction. In the 4-terminal device, where the average
electrochemical potential is equal to that of the superconductors
($\mu_s = 0$), there is no voltage drop across either NS junction. The
presence of absence of these voltage drops does not (to lowest order)
affect the supercurrent flow.

\subsection{Short Junction}
\indent

Figure~\ref{fig:shortJJ} shows the terminal characteristics of a short
($L \ll \xi_0$, where $L$ is the length of the normal region and
$\xi_0$ the coherence length of the superconductors) Josephson
junction.  The gate couplings in Fig.~\ref{fig:shortJJ} are weak, with
$T_1 = T_2 = 1 \% \equiv \epsilon$.  We fix the superconducting phase
difference at $\phi = .6 \pi$. The differential conductance along the
gate shows a resonance whenever the gate voltage crosses a new Andreev
level in Fig.~\ref{fig:shortJJ}(a), a tool which can be used to
perform `Andreev level spectroscopy'. The gate current versus voltage
in Fig.~\ref{fig:shortJJ}(c) is simply the integral of
Fig.~\ref{fig:shortJJ}(a), showing steps in the gate current instead
of peaks in differential conductance. The Josephson current versus
gate voltage in Fig.~\ref{fig:shortJJ}(b) switches abruptly to zero
whenever a new Andreev level is populated. Reducing transmission $T$
across the normal region from $100 \%$ to $12 \%$ in
Fig.~\ref{fig:shortJJ}(b) suppresses the Josephson current and pushes
the Andreev level closer to the gap edge, pushing the switching
voltage close to $eV_G \simeq \Delta$.

The Josephson current versus gate current is shown in
Fig.~\ref{fig:shortJJ}(d).  Both the Josephson current switching in
Fig.~\ref{fig:shortJJ}(b) and the increase in the gate current in
Fig.~\ref{fig:shortJJ}(c) occur over over the same range of voltage,
namely the width (in energy) of the Andreev level.  The Josephson
current therefore decreases quite slowly and gradually with increasing
gate current. The fractional occupation of the single Andreev level in
this junction ($L \ll \xi_0$) also increases slowly from $f_B =0$ to
$f_B = 1$ during the switching event. Contrast the sharp switching
characteristics in the Josephson current versus gate voltage from
Fig.~\ref{fig:shortJJ}(b) (roughly a square wave) to the relatively
slow change of Josephson current versus gate current in
Fig.~\ref{fig:shortJJ}(d) (approximately triangular).

\begin{figure}[thb]
\threefig{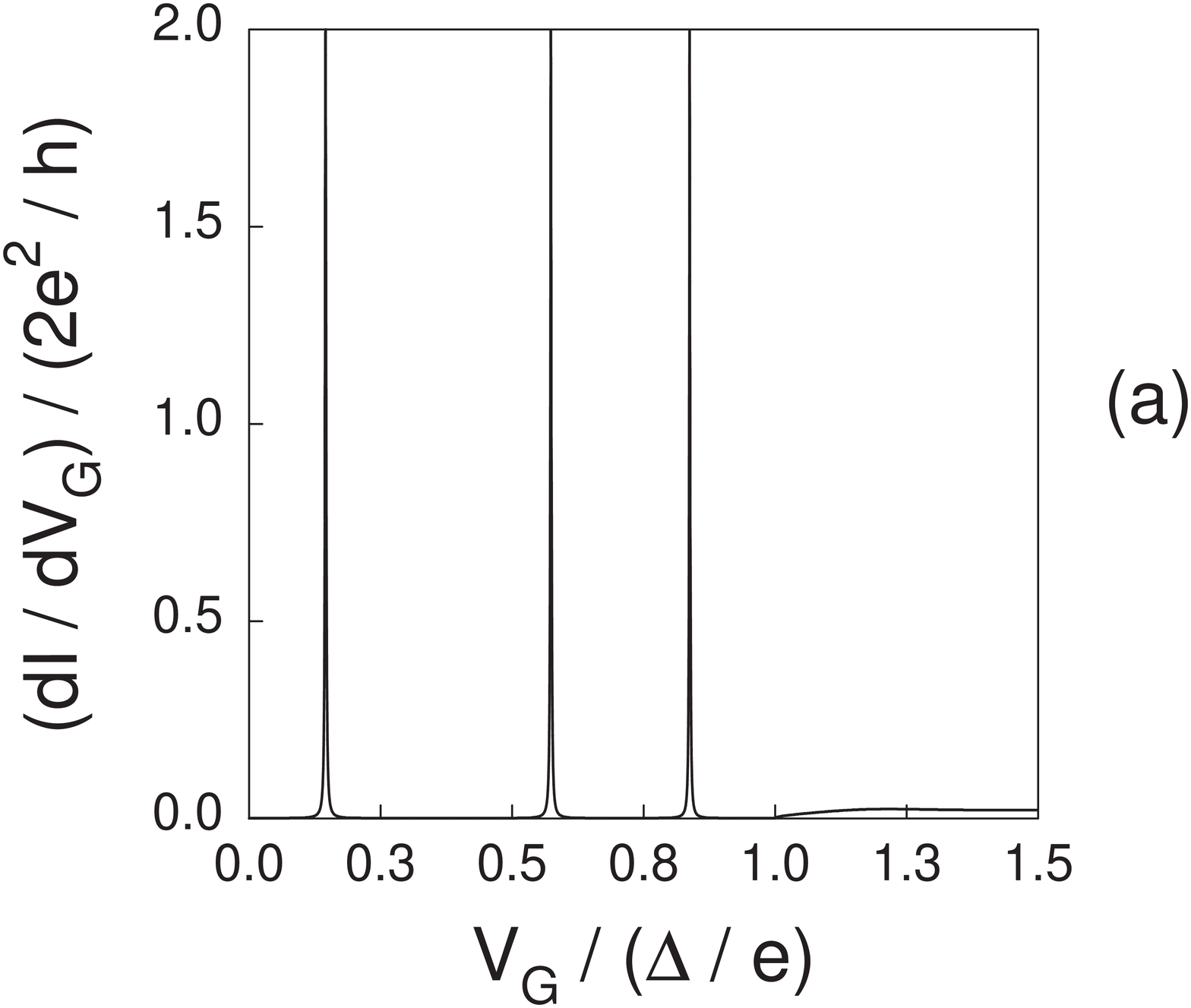}{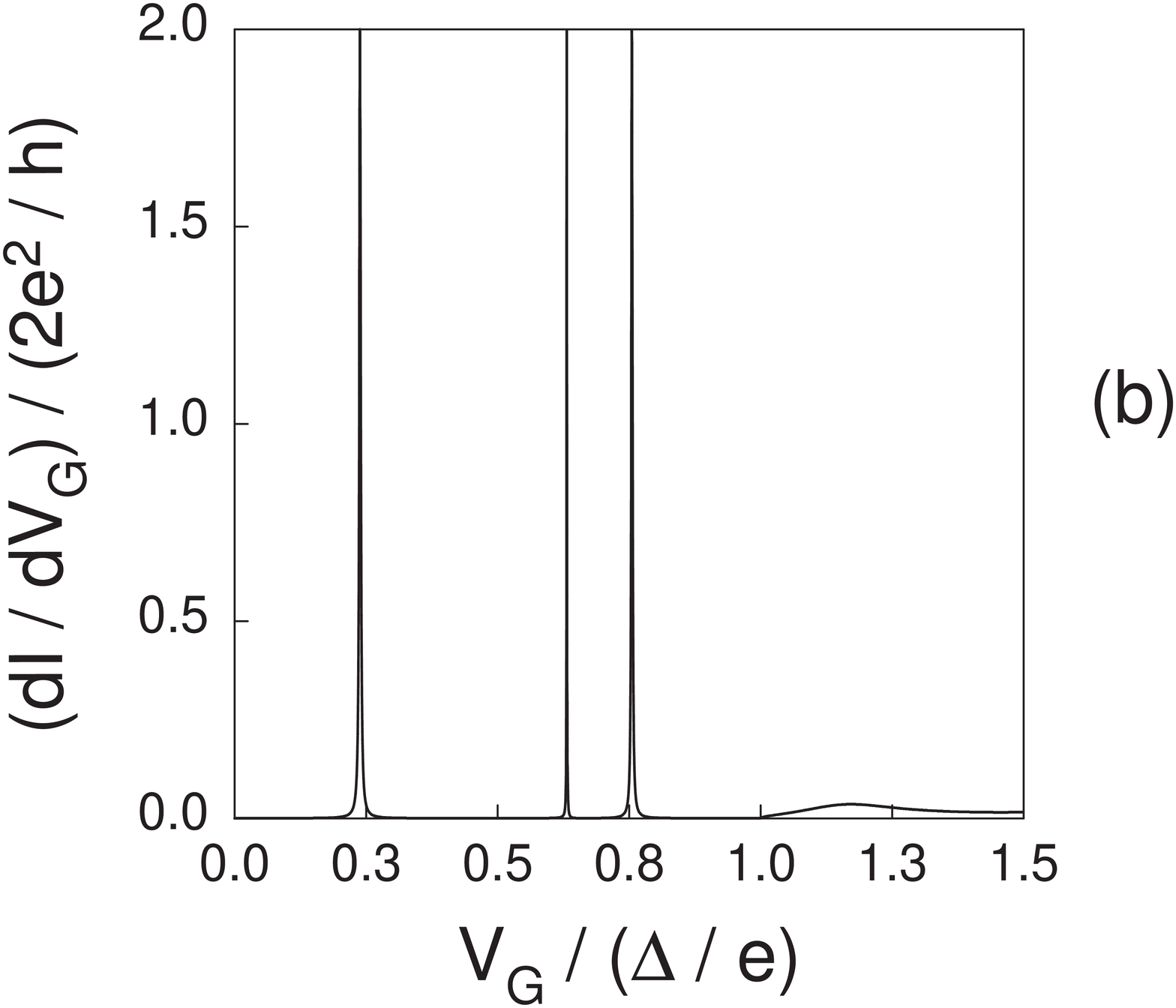}{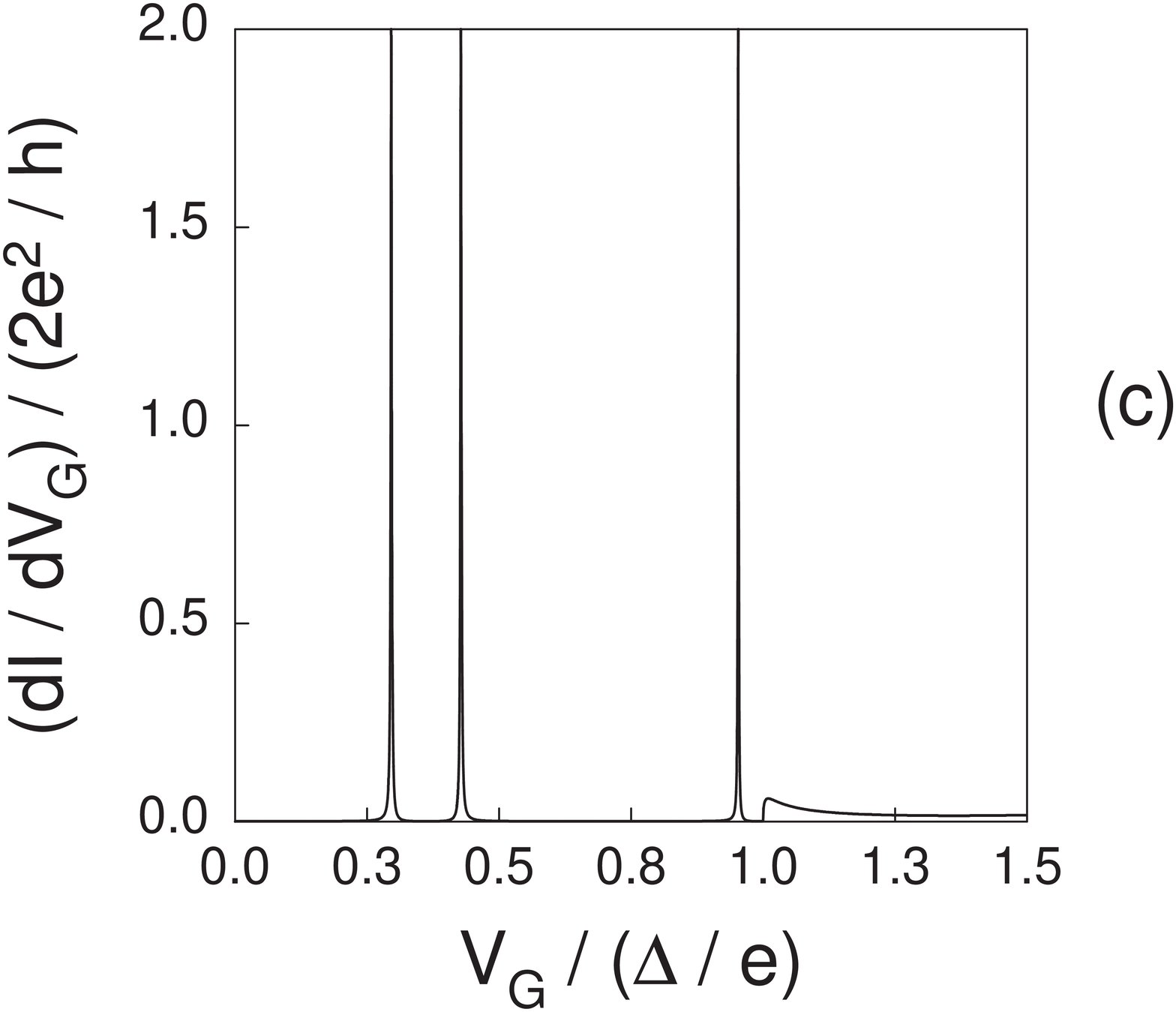}{5.8}
\caption{The differential conductance along the gate versus gate
voltage shows more Andreev levels for the long ($L \gg \xi_0$)
Josephson junction. The long junction is either (a) ballistic, 
(b) an `asymmetric' tunnel junction, or (c) a
`symmetric' tunnel junction.}
\label{fig:longdidv}
\end{figure} 

\begin{figure}[thb]
\threefig{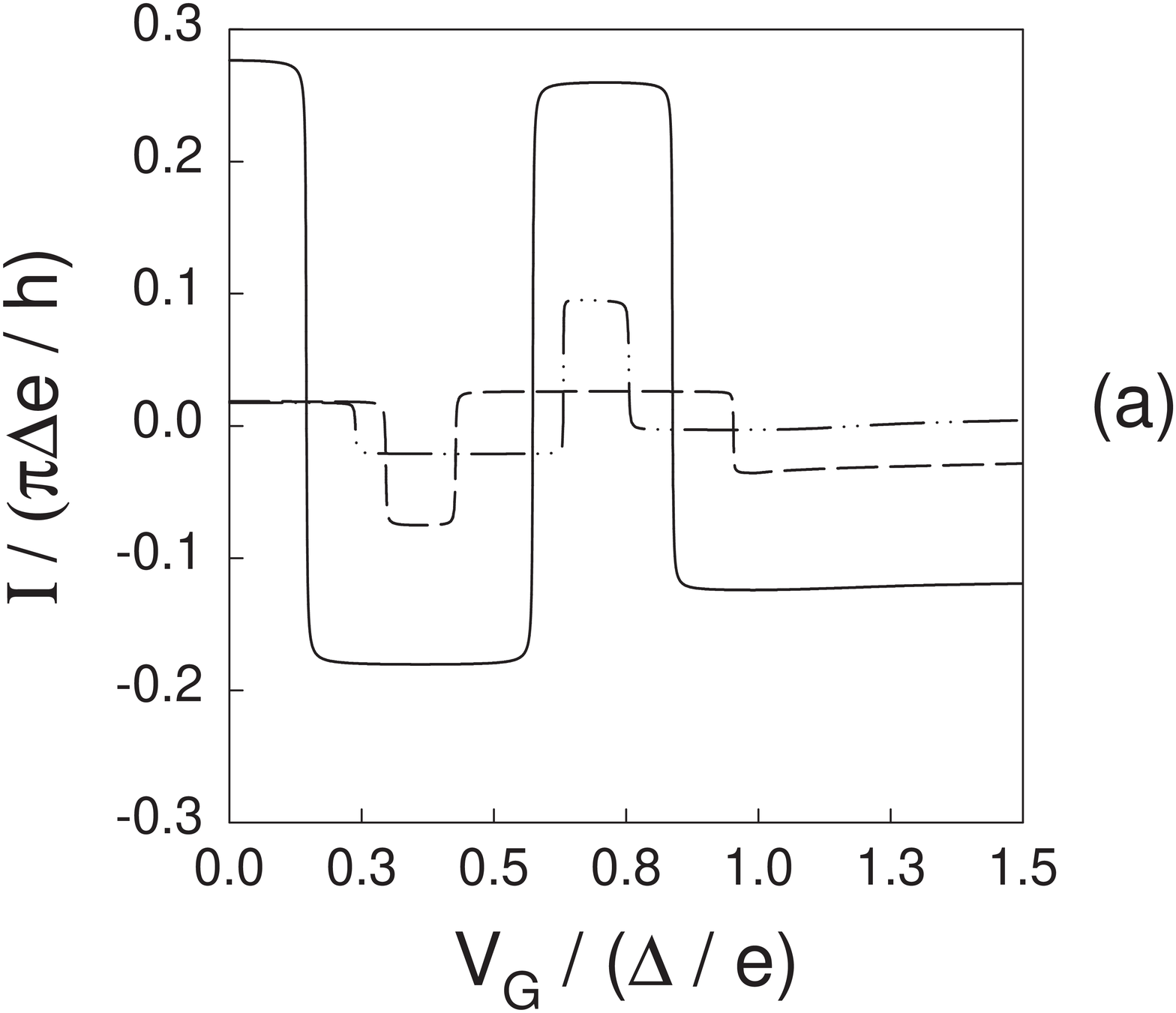}{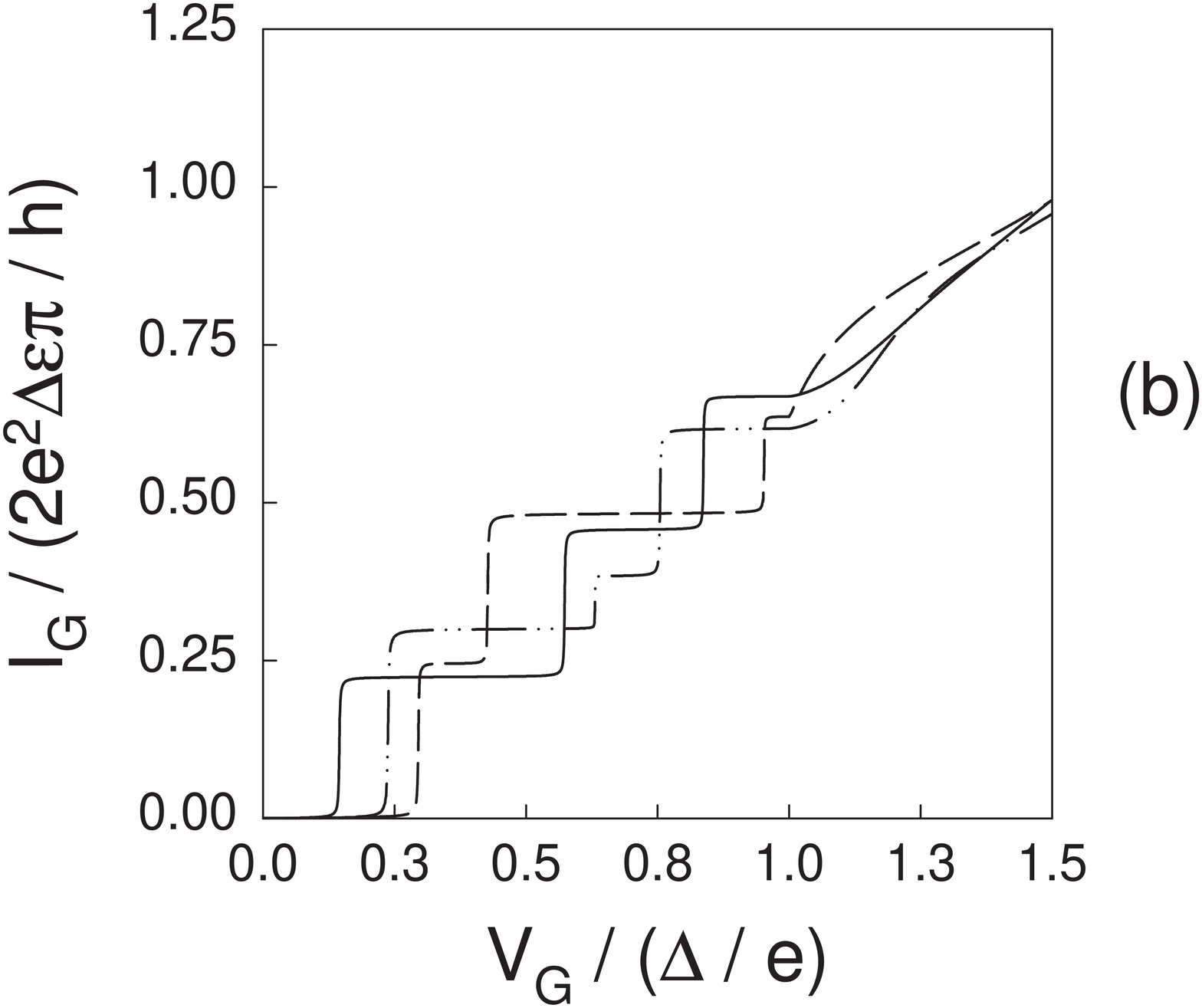}{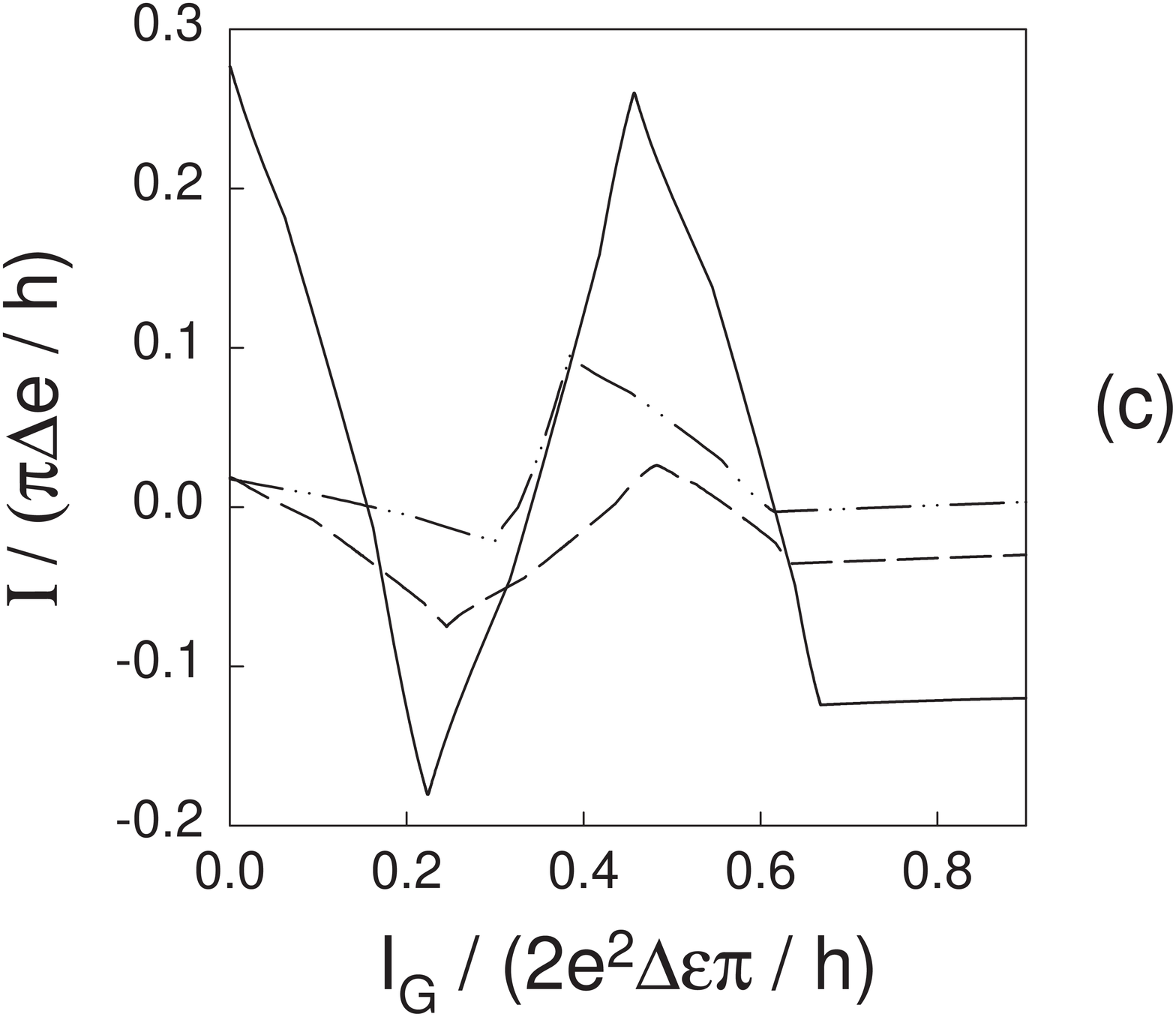}{5.8}
\caption{For a long Josephson junction, which is either
ballistic (solid), a `symmetric' tunnel junction (dashed), 
or an `asymmetric' tunnel junction (dot-dashed), we show 
(a) the Josephson current versus gate voltage, 
(b) the gate current versus gate voltage, and
(c) the Josephson current versus gate current.}
\label{fig:longcur}
\end{figure} 

\subsection{Long Junction}
\indent

Figures~\ref{fig:longdidv} and \ref{fig:longcur} show the same
terminal characteristics for a long ($L \gg \xi_0$) Josephson
junction. In long SNS junctions the position of the tunnel barrier has
some influence on the size of the Josephson
current~\cite{bagwell,ilhan,wendin}.  We therefore consider a
`ballistic' junction with $T=1$, an `asymmetric' junction with the
tunnel barrier $1/5$ of the distance across the normal region ($a =
L/5$ and $T=12 \%$), and a `symmetric' junction with the tunnel
barrier in the middle of the normal region ($a = L/2$ and $T=12 \%$).
The gate is again only weakly coupled to the Josephson junction, with
$T_1 = T_2 = 1 \% \equiv \epsilon$. We again fix $\phi = .6 \pi$.
This particular long junction has three Andreev levels in the energy
range within $\Delta$ of the Fermi level (since $L = 6.6 \xi_0$), as
can be seen from the differential conductance along the gate versus
gate voltage in Fig.~\ref{fig:longdidv}. Only small differences appear
in the differential conductance along the gate between the different
types of long junctions in Fig.~\ref{fig:longdidv}.

Figure~\ref{fig:longcur} shows (a) the Josephson current versus gate
voltage, (b) the gate current versus gate voltage (the integral of
Fig.~\ref{fig:longdidv}), and (c) the Josephson current versus gate
current for the three different types of long Josephson junctions.
The Josephson current in Figs.~\ref{fig:longcur}(a) and (c) changes
sign whenever the gate voltage is such that a new Andreev level
becomes populated or depopulated~\cite{vwees,chang} (sometimes called
a `$\pi$-phase shift'~\cite{volkov,wilhelm}). A sign change of the
Josephson current only occurs in long junctions, as filling or
emptying the single Andreev level in a short junction forces the
Josephson current to zero. The Josephson current remaining in
Fig.~\ref{fig:longcur}(a) and (c) when all the Andreev levels are
filled ($eV_G \ge \Delta$), is the `continuum' current $I_C$.

The Josephson current switches abruptly with gate voltage in
Fig.~\ref{fig:longcur}(a), but only gradually with gate current in
Fig.~\ref{fig:longcur}(c). The with in voltage over which these
currents change is again the Andreev energy level width.  Andreev
level is filling changes the gate current and Josephson current
simultaneously, leading again to only slow variation of Josephson
current with gate current in Fig.~\ref{fig:longcur}(c). The same
reasoning explains why the shape of Fig.~\ref{fig:longcur}(a) is
roughly a square wave, while Fig.~\ref{fig:longcur}(c) is roughly
triangular.

The ballistic junction (solid lines) always has the largest Josephson
current, as seen in Fig.~\ref{fig:longcur}(a) and (c). The symmetric
junction (dashed lines) displays the so-called `giant' Josephson
current~\cite{wendin,ilhan} as the switching event near $eV_G \simeq
.35 \Delta$ in Fig.~\ref{fig:longcur}(a), but the asymmetric junction
can also display switching currents of the same magnitude. One indeed
expects the symmetric tunnel junction to have large switching currents
compared to its Josephson current~\cite{wendin}, but the asymmetric
junction can have these as well when the Andreev gaps are
small~\cite{ilhan}.

Since Ref.~\onlinecite{morpurgo} uses metals to fabricate the
Josephson junction, the quasi-particle trajectories forming the bound
Andreev levels will not be the simple types of scattering trajecties
assumed in Refs.~\cite{chang,ilhan}, but will be altered by impurity
scattering inside the Josephson junction.~\cite{volkov,wilhelm} In
such dirty junctions at very low temperature another energy scale
becomes important (the Thouless energy, set by the time for an
electron to diffuse across the Josephson junction). This extremely low
energy scale does not increase as the superconductor improves, and is
not essential to understanding the Josephson current switching (or
`$\pi$-phase shifting'). The Josephson current changes sign in long
junctions whenever a new Andreev level is populated or depopulated.
Populating a new Andreev level in the short junction simply forces the
Josephson current to switch off.

\subsection{Maximizing Current Gain}
\indent

We now consider how to maximize the change of Josephson current over
the change in gate current while the Josephson current switches.  For
simplicity we consider only the ballistic junction, $T=1$. Following
Ref.~\onlinecite{chang}, the change in Josephson current $(\Delta I)$
is
\begin{equation}
(\Delta I) = \frac{e v_F}{L+2\xi(E_n)}
\end{equation}
with $\xi(E_n)$ the energy dependent coherence length. The change
in gate current near a resonance is
\begin{equation}
(\Delta I_G) = \epsilon \frac{e v_F}{L+2\xi(E_n)} \; , 
\end{equation}
where $T_1 = T_2 \equiv \epsilon$ is the gate coupling. The ratio
of Josephson to control current is therefore
\begin{equation}
\frac{(\Delta I)}{(\Delta I_G)} = \frac{1}{\epsilon} \; .
\end{equation}
If we could take the gate coupling to zero, the current gain would
become infinite. However, inelastic or phase breaking scattering sets
a lower limit on the gate coupling needed for the gate control to
affect the Josephson current.~\cite{vwees} If the inelastic scattering
time is $\tau_{\phi}$, we have~\cite{chang}
\begin{equation}
\epsilon_{\rm min} = \frac{L+2 \xi_0}{v_F \tau_{\phi}} \; .
\end{equation}
The maximum current gain then becomes
\begin{equation}
\left. \frac{(\Delta I)}{(\Delta I_G)} \right|_{\rm max} = 
\frac{v_F \tau_{\phi}}{L+2 \xi_0} \to
\frac{\Delta \tau_{\phi}}{\hbar} \; , 
\end{equation}
where the arrow indicates the short junction limit. The maximum
current gain in the ballistic short junction limit is therefore set by
the ratio of the superconducting energy gap to the level broadening
induced by inelastic or phase breaking scattering. The device of
Ref.~\onlinecite{schalpers}, in the short, ballistic junction limit,
shows a current gain of about 20 for small gate currents.

\section{Conclusions}
\indent

We have shown the dependence of the supercurrent flow in a three or
four terminal Josephson junction on the current passing through a
normal metal gate, motivated by two recent
experiments.~\cite{morpurgo,schalpers} When the gates or gates are
weakly coupled to the Josephson junction, desirable for better
isolation of gate (control) current from the Josephson current, there
is little (or no) difference between the terminal characteristics of
the three versus four terminal geometries. The Josephson current in
either geometry is a slowly varying function of the gate current, even
for the idealized one-dimensional model we have considered in
this paper. The Josephson current varies much more rapidly as a
function of the gate voltage. Measuring the differential conductance
along the gate would also permit direct observation (spectroscopy) of
the current carrying energy levels inside the superconducting gap.

In both the transistor fabricated using metals~\cite{morpurgo} and in
compound semiconductors~\cite{schalpers}, the gates are not well
isolated from the device channel.  Instead of the tunnel barriers
drawn in Fig.~\ref{fig:geom}, only impurity scattering inside the Au
metal~\cite{morpurgo} or two-dimensional electron
gas~\cite{schalpers}, along with possible contact resistances, provide
partial gate isolation. Since quasi-particles which are normally
trapped inside the Josephson junction can now escape by leaking into
the gate, the bound Andreev levels in either device will therefore be
very broad. Broad Andreev levels not only broaden the Josephson
current switching, but produces gate currents in experiments which are
comparable to the supercurrents. As pointed out in
Ref.~\onlinecite{schalpers}, The two dimensional (planar) geometry of
the devices used in the two experiments also introduce a type of
geometrical broadening~\cite{orlando} which needs to be studied in
this context.

\section{Acknowledgements}
\indent

We thank T. Sch\"{a}lpers for communicating his experimental results
prior to publication.  We gratefully acknowledge support from the
MRSEC of the National Science Foundation under grant No. DMR-9400415
(PFB).

\end{document}